\documentstyle[12pt]{article}
\begin{document}
\title{BLACK HOLE INFORMATION \thanks
{Alberta-Thy-23-93, hep-th/9305040.
Invited review lecture published in {\em Proceedings of the
5th Canadian Conference on General Relativity and Relativistic
Astrophysics,
University of Waterloo, 13--15 May, 1993}, edited by R. B. Mann and
R. G. McLenaghan (World Scientific, Singapore, 1994), pp. 1-41.}}
\author{ Don N. Page\\
CIAR Cosmology Program\\
Theoretical Physics Institute\\
Department of Physics\\University of Alberta\\
Edmonton, Alberta\\Canada T6G 2J1\\
Internet:  don@page.phys.ualberta.ca}
\date{(1993 May 10, revised July 31)}
\maketitle
\large
\begin{abstract}
\baselineskip 16pt

     Hawking's 1974 calculation of thermal emission
from a classical black hole led to his 1976 proposal
that information may be lost from our universe as a
pure quantum state collapses gravitationally into a
black hole, which then evaporates completely into
a mixed state of thermal radiation.  Another possibility
is that the information is not lost, but is stored in a
remnant of the evaporating black hole.  A third idea
is that the information comes out in nonthermal
correlations within the Hawking radiation, which
would be expected to occur at too slow a rate,
or be too spread out, to be revealed by any
nonperturbative calculation.
\\
\\
\end{abstract}
\normalsize
\pagebreak
\baselineskip 17pt

\vspace{5 mm}
{\bf 1.  Hawking's Proposed Loss of Information}
\vspace{5 mm}

     Hawking's 1974 calculation \cite{Haw74,Haw75} of the emission
from a
stationary classical black hole was soon shown to give uncorrelated
thermal
emission in each mode [3-5].  If a semiclassical
approximation were used so that the black hole shrinks in a
quasistationary way
during the evaporation, one would still expect nearly thermal
emission with
high entropy, though even this calculation (without quantizing the
geometry)
has not been done precisely in four dimensions.  As a result of these
calculations and
expectations, Hawking argued \cite{Haw76} that the semiclassical
approximation
should be good until the black hole shrunk near the Planck mass, and
then there
would not be enough energy left for the information that collapsed
into the
black hole to come back out.  Thus a pure quantum state that
underwent
gravitational collapse into a black hole that subsequently evaporated
away
would end up as a mixed quantum state of Hawking radiation.  Hawking
proposed
that the process would be described by a superscattering operator
$\$$ that
would take initial density matrices into final density matrices in a
nonunitary
way, generically increasing the fine-grained entropy
$S\equiv-Tr(\rho\ln\rho)$:
     	\begin{equation}
	\rho^{final}_{ab}=\$_{ab}^{cd}\rho^{initial}_{cd},
	\end{equation}
with a sum over the repeated indices $c$ and $d$,
where $\$_{ab}^{cd}$ would not have the usual unitary form
$S_a^c\bar{S}_b^d$
in terms of a unitary $S$ matrix.

     This process would correspond to the loss of information in the
sense that
from an initial pure state, one could not predict any single final
pure state
with certainty.  A pure state may be represented by a normalized
vector
$|\psi\rangle$ in the Hilbert space of the quantum system (here, the
universe,
or at least our connected component of it).  A pure state can also be
represented by a statistical state or density matrix (in general, a
positive-semidefinite Hermitian unit-trace operator or matrix acting
on the
vectors in the Hilbert space) which when pure has the form
$\rho=|\psi\rangle\langle\psi|$.  Hence in this pure case $\rho$ is a
rank-one
projection operator with $\rho^2=\rho$ and therefore with
$Tr\rho^2=1$
and entropy $S=0$ as well as the general normalization (unit-trace)
requirement $Tr\rho=1$.
A pure state may be contrasted with a mixed state, which cannot be
represented
by a single vector in the Hilbert space, but which can be represented
by a
statistical state or density matrix $\rho=\sum_n p_n
|\psi_n\rangle\langle\psi_n| \neq \rho^2$ with more than one nonzero
eigenvalue
$p_n$, each of which can be interpreted to be the probability of
measuring the
mixed state to be in one of the corresponding orthonormal pure states
 $|\psi_n\rangle$.  (It can be misleading, e.g., in the EPR
`paradox,'
 to say that a system in a mixed state
{\it is} actually in one of its component pure states
$|\psi_n\rangle$ with
probability $p_n$, because the system might instead actually be
correlated with
another system in a composite system, and the composite system as a
whole could
even be in a pure state.)  By the positive-semidefinite and
unit-trace
properties of general density matrices, the $p_n$'s are nonnegative
and add up
to unity, and one can readily see that a mixed state has $Tr\rho^2 <
1$ and
$S\equiv -Tr(\rho\ln\rho)>0$.

     If one makes a complete measurement of a system, by which I mean
the
measurement of a nondegenerate observable (represented by some
Hermitian
operator with totally nondegenerate eigenvalues), a pure state is the
only kind
that can give a definite result with certainty (unit probability).  A
pure
state, and only a pure state, has the property that one can predict
with
certainty the result of {\it some} complete measurement of the
system.  For one
indeed to be able to predict with certainty, one needs the measured
observable
to have one of its eigenvectors proportional to the Hilbert space
vector
$|\psi\rangle$ representing the pure state, so of course not every
observable
will give a uniquely predictable result.  (This is a manifestation of
the
quantum uncertainty that applies even to pure states.)  However, for
any pure
state, there do exist nondegenerate observables (in the sense of
Hermitian
operators but not, in general, in the sense of what is experimentally
and
practically possible) which would give definite results.

     For example, suppose one had
access to a sufficiently large ensemble of identical systems in the
same
(initially
unknown) pure state and could in principle measure enough
observables.  The
first observable one randomly chose to measure would generally not
give
definite results (i.e., one would get different individual results
when one
measured different members of the ensemble with it).  Nevertheless,
one could
eventually find some observable which would always give the same
definite
result when applied to members of the ensemble.  (This would not be a
unique
observable, since only its eigenvector proportional to the pure-state
vector
would be uniquely determined, up to a complex multiplicative
constant.
Furthermore, this member of the preferred class of observables would
only
be determined to some finite accuracy if
only a finite sequence of measurements were made to find a finitely
good
empirical approximation to one of these preferred observables.)

    It is in this sense that
one can say that complete information (the maximum allowed by quantum
mechanics) exists for a system in a pure state.

     On the other hand, for a mixed or impure state there is no
nondegerate
observable that would have a unique value with unit probability.  In
other
words, one cannot predict with certainty the result of {\it any}
complete
measurement of a system in a mixed state.  In this sense one says
that mixed
states have less than maximal information.  Of course, in another
sense one
could say that the actual density matrix gives all of the information
possible
about the state of a system (at least if one considers the system by
itself,
ignoring any quantum correlations it may have with other systems).
In this
latter sense, if one knows the actual density matrix (not just an
estimate
based on additional uncertainty about the actual statistical state of
the
system), one has complete information about the system in whatever
statistical
state it is actually in.  However, it is in the former sense, of how
much
information is {\it possible} about the system (which depends on the
actual
statistical state rather than on how well that is known), that one
says that
the evolution of a pure state to a mixed state by Hawking's proposed
Eq. 1 for
black hole formation and evaporation would be a loss of information.
A measure
of the possible amount of information in a system with Hilbert-space
dimension
$m$ and statistical state $\rho$ is
     	\begin{equation}
	I = S_{max} - S=\ln m + Tr(\rho\ln\rho).
	\end{equation}

     This loss of information does {\it not} necessarily mean that a
knowledge
of the final mixed state would be insufficient to reconstruct the
initial
state.  (The final mixed state could in principle be learned to
arbitrarily
high statistical accuracy from the results of repeated measurements
of a
sufficiently large set of observables if one had a sufficiently large
ensemble
of systems with that same mixed state.)  To illustrate this claim and
the
points made above, consider a spin-1/2 system, so that the indices
${a,b,c,d}$
range from 1 to 2.  The general statistical state of the system can
be written
as
     	\begin{equation}
	\rho=\rho_{11}|\uparrow\rangle\langle\uparrow|
	+\rho_{12}|\uparrow\rangle\langle\downarrow|
	+\rho_{21}|\downarrow\rangle\langle\uparrow|
	+\rho_{22}|\downarrow\rangle\langle\downarrow|,
	\end{equation}
where $\rho_{11}$ and $\rho_{22}$ must be nonnegative real numbers
that sum to
unity and $\rho_{12}$ and $\rho_{21}$ must be complex conjugates with
$\rho_{12}\rho_{21} \leq \rho_{11}\rho_{22}$ in order that $\rho$ be
a
positive-semidefinite Hermitian unit-trace operator.  This
statistical state
can be characterized by the polarization vector ${\bf P}$ with
Cartesian
components
$(\rho_{12}+\rho_{21},i\rho_{12}-i\rho_{21},\rho_{11}-\rho_{22})$
and gives a probability $(1+P)/2$, where $P$ is the magnitude of
${\bf P}$, of
finding the spin in the direction of ${\bf P}$.

    Now an example of a superscattering matrix for this system is
     	\begin{equation}
	\$_{ab}^{cd} = \lambda\delta_a^c\delta_b^d +
\frac{1}{2}(1-\lambda)\delta_{ab}\delta^{cd},
	\end{equation}
 which simply multiplies the polarization vector by the real number
$\lambda$,
where $0\leq\lambda \leq 1$.  If $\lambda=1$, this gives the trivial
unitary
transformation which takes a pure state (one with $\rho_{12}\rho_{21}
=
\rho_{11}\rho_{22}$ or $P=1$, which has the spin pointing in the
direction of
the unit vector ${\bf P}$ with unit probability) to the same pure
state.
However, if $\lambda<1$, $\$$ takes a pure state to a mixed state.
But unless
$\lambda=0$, a determination of the final state (e.g., by a
successive
determination of the expectation values of the three Cartesian
components of
the spin for a large ensemble of systems with this identical final
state) would
readily give the initial state, simply by dividing the polarization
vector by
$\lambda$.  (Of course, I am always assuming that the law of
evolution, in this
case parametrized by $\lambda$, has already been determined, as it
could have
been by measuring the results of the evolution of an ensemble of
systems in
different initial states.)  Only in the special case $\lambda=0$ does
the final
statistical state not uniquely determine the initial statistical
state.

     Although the simple case just given is quite special, the
property appears
to be generic, that a superscattering matrix is invertible within the
restricted set
of density matrices comprising its range, if the dimensions of both
the initial
and final Hilbert spaces are the same finite integer $m$.  That is,
for the set
of $(m^2-1)^2$ real parameters defining the generic superscattering
matrix,
all but a set of measure zero, given by one or more hypersurfaces of
codimension one, or dimension $m^2(m^2-2)$, in the
$(m^2-1)^2$-dimensional
space of all the parameters, gives an invertible $\$$.  Of course, if
$\$$
increases the entropy $S$ and decreases $Tr\rho^2$, there are
hypothetical
positive-semidefinite final density matrices (e.g., pure states) that
have no
pre-images by $\$$ in the space of positive-semidefinite initial
density
matrices.
(The inverse of $\$$ would map them to matrices with one or more
negative
eigenvalues.)  However, the generic $\$$ would have an inverse within
the
smaller set of final density matrices given by the range of $\$$
acting on the
set of general positive-semidefinite initial density matrices of the
same
finite
dimension.  On the other hand, if the dimension of the final Hilbert
space
were smaller than that of the initial one (a rather violent violation
of
$CPT$),
no superscattering matrix could be invertible.

     Thus
one might say that a nonunitary superscattering operator does not
generally
lead to an absolute loss of information about the initial state,
assuming
the Hilbert spaces stay the same finite dimensions. However, it does
lead to a degradation of partial information in the sense that the
empirical
accuracy of the measured state is reduced in extrapolating back
(e.g., in
dividing the final approximately-determined polarization vector by
$\lambda$).
Nevertheless, I shall continue to follow the usual convention of
defining a
loss of information as an increase in the fine-grained entropy
$S\equiv-Tr(\rho\ln\rho)$ of the complete system.

     This loss of information proposed by Hawking would be a new
feature
of quantum gravity, not seen in other quantum field theories in a
fixed
globally hyperbolic spacetime.  Gibbons \cite{Gib77} speculated that
it might be related to the indefiniteness of the Einstein action of
general
relativity for positive-definite metrics (Riemannian or `Euclidean'
as
opposed to pseudo-Riemannian or Lorentzian).

\vspace{5 mm}
{\bf 2.  Initial Objections and Alternatives to Hawking's Proposal}
\vspace{5 mm}

     To the best of my knowledge, the first objection to Hawking's
proposal
of a loss of information was made by the referee \cite{Hawpri}, who
apparently forced Hawking to change the title from ``Breakdown of
Physics \ldots '' to ``Breakdown of Predictability \ldots .''
However, I am
not aware of whatever detailed objections he may have given.  So far
as I
know now, the first published objection was given by Zel'dovich
\cite{Zel77b},
who asked whether Hawking's ``very radical'' conclusion is
``connected
with the fact that he considers a macroscopic black hole?\ldots
Could not
this new and greater indeterminacy arise as a result of this
macroscopic
and semiclassical treatment of the situation?\ldots  Must one not
treat
the emission of a black hole at the quantum level?  Can one not, and
should one not formulate the theory with black holes in such a way
that
additional indeterminacy and incoherence do not arise?''

     Unaware of Zel'dovich's objection, I independently objected to
Hawking's proposed nonunitary evolution Eq. (1) on the grounds that
it
is not $CPT$ invariant, since it takes
pure states to mixed states, but it does not give the $CPT$-reversed
process of
mixed states going to pure states \cite{Pag80}.  After making that
new
observation, I raised the same objection that Zel'dovich had
\cite{Zel77b},
that Hawking's proposal was based on the semiclassical approximation
(SCA).
I explicitly showed how the SCA
would be expected to break down in a noticeable way
from fluctuations of the black hole momentum, long before the black
hole shrunk
to the Planck size, if momentum is conserved in detail and not just
on average.
Although this particular fluctuation effect by itself would not help
restore
any of the information Hawking believed would be lost, it did at
least
illustrate the uncertainty of relying on the SCA for
the question of whether the results of black hole evaporation are
predictable
in the sense that a pure quantum state would be.  Therefore, I listed
a number
of open possibilities that occurred to me at that time, which it may
be useful
for me to rewrite and comment on as I see them today:

     (A) Evolution by an $S$ matrix.  This would say that in a
quantum theory
of everything, including the gravity of the black hole, a pure state
of the
complete system would always go to a pure state, and no information
would be
lost.  At the time, I wrote that ``in the absence of further
information, it
would seem most productive to pursue the most conservative
possibility (A).''
Today I might be somewhat inclined to delete the word ``most,'' but I
have not
yet seen any strong evidence that (A) does not remain an open
possibility, for
reasons I shall partially discuss below.  In some sense it would be
the
simplest possibility, though I must admit I still have little idea
how it might
be actually realized and yet be consistent with what we think we know
about
gravity.  More recent arguments for this viewpoint include
[10-49], as will be discussed below.

     (B) Evolution by a $CPT$-noninvariant superscattering matrix.
This was
Hawking's proposal \cite{Haw76}.  It seemed undesirable to me to
believe in a
violation of $CPT$ invariance, despite the previous arguments of
Penrose
\cite{Pen79}, but Hawking, though not one to believe Penrose's
arguments on
this, concurred \cite{Haw80pri} with Wald [52-54] that it
would be enough to have $CPT$ in the weak form of $CPT$-invariant
transition
probabilities
     	\begin{equation}
	p(c\rightarrow a)\equiv\$_{aa}^{cc}=p(\Theta
a\rightarrow\Theta c)
	\end{equation}
between an initial pure state $c$ and a final pure state $a$
(no sum on these repeated indices), by using Eq. 1 as
an intermediate tool but not interpreting the final density matrix
given there
as literally the actual final state of the system.  I have found this
hard to
swallow in my na\"{\i}vely realist view of density matrices as being
the more
basic objects, and of probabilities as being derived from them,
rather than the
other way around.  However, Hawking's argument \cite{Haw82}  that one
should interpret Eq. 1 as merely an intermediate
tool for calculating conditional probabilities (given a measurement
of a
particular initial pure state, what is the conditional probability of
measuring
a particular final pure state?) now makes more sense to me
\cite{Pag93notime}.  Then the asymmetry may indeed be more in the
{\it
conditional} nature of the probability than in any {\it time}
asymmetry (e.g.,
$CPT$ noninvariance).

     (C) Evolution backward in time by a $CPT$-noninvariant
superscattering
matrix.  I threw this in as a counterpoint to (B) and wrote, ``Then
the past
would be predictable from the future, but the future would not be
predictable
from the past.  This possibility would suit historians better than
physicists.''
 I suppose that if the superscattering matrix were merely used to
calculate
conditional probabilities, and if (B) were applicable when the
condition were
to the past of the result whose conditional probability were to be
calculated,
then (C) would apply whenever the result were to the past of the
condition.
This indeed appears to be roughly the case when historians ask,
``What is the
probability that $x$ happened in the past, given our present
records?''

     (D) Evolution by a superscattering matrix which is not of the
form Hawking
proposed, i.e., not
     	\begin{equation}
	\$_{ab}^{cd} = S_{aA}^c \bar{S}_{bA}^d,
	\end{equation}
where $A$ denotes an orthonormal basis of states (to be summed over)
of the
``Hilbert space of all possible data on the hidden surface'' which is
to
surround ``either singularities (as in the Schwarzschild solution) or
`wormholes' leading to other space-time regions about which the
observer has no
knowledge (as in the Reissner-Nordstr\"{o}m or other solutions)''
\cite{Haw76}.
$S_{aA}^c$ would be an $S$-matrix from a state $c$ on an initial
surface
(before a black hole formed) to a state $aA$ with $a$ on a final
surface (after
the black hole evaporated) and $A$ on the hidden surface.  In my
Letter I
explicitly assumed  an $\$$ with this form in (B), and also in (C),
except with
``initial'' and ``final'' reversed in (C).  I didn't have any
motivation for
considering other forms of a possible superscattering matrix, which
would be of
no help in avoiding violating $CPT$ invariance in my strong sense.
However,
it now appears to be necessary if the initial and final Hilbert
spaces have the
same finite dimension (see below).

     (E) Evolution of density matrices deterministically but
nonlinearly.
I don't see any clear motivation for this, but why not leave it on
the table as
an open possibility?  Nonlinear generalizations of the quantum
mechanical
evolution of pure states have been considered [57-60],
but I am not aware of much discussion of nonlinear evolution of
density
matrices that do not keep pure states pure \cite{PreSB}.
The apparent linearity of quantum mechanics
seems to me to be the main reason why we do not notice any influence
from
other Everett worlds, which surely must exist unless quantum
mechanics is
modified in a very particularly nonlinear way (e.g., by the collapse
of the
wavefunction somehow very precisely into just the quasiclassical
components
we observe).  I would think that any proposed nonlinearities of
quantum
mechanics would be very strongly limited by our nonobservance of such
other-worldly effects.

     (F) Evolution in which black holes or naked singularities form
but do not
disappear.  These would now be called remnants or various other terms
[62-65], but this term is also taken to mean other
massive
objects that I did not think to consider \cite{Gid92}, and also
objects that
can decay unitarily after a very long time \cite{ACN,Pre92}, which I
would
have
counted as an intermediate state in possibility (A) but did not
consider
explicitly.  As far as the absolutely stable remnants go, I
uncritically
accepted Hawking's argument that ``Because black holes can form when
there was
no black hole present beforehand, $CPT$ implies that they must also
be able to
evaporate completely; they cannot stabilize at the Planck mass, as
has been
suggested by some observers'' \cite{Haw76,HawHay}.  However,
Horowitz \cite{Hor93pri}
led me to realize that the only requirement from $CPT$ is that a
$CPT$-reversed
remnant should be able to combine with $CPT$-reversed Hawking
radiation to form
a large $CPT$-reversed black hole (i.e., a white hole) which can
convert into
the $CPT$ reverse of whatever collapsed to form the original black
hole; if
there is no $CPT$-reversed Hawking radiation impinging on the
$CPT$-reversed
remnant, it can be absolutely stable and yet be consistent with $CPT$
invariance.

     It would seem that there could be several possibilities for a
set of
absolutely
stable (when isolated) remnants resulting from black hole evaporation
that
would be consistent with $CPT$:  (1) The set could be empty.  Then
stable
remnants would not exist.  (2) The set could be nonempty and include
the
$CPT$ reverse of every element of the set (or at least a quantum
superposition
thereof).  Then each remnant could in principle be made to go away by
combining it with the $CPT$ reverse of the Hawking radiation that
accompanies the formation of the $CPT$-reversed remnant.  (3) The set
of remnants could be nonempty but distinct from the $CPT$-reversed
set of anti-remnants.  (For example, they could be cornucopia
geometries
[62-65]
with internal regions that are expanding toward internal future null
or
timelike
infinities, whereas anti-remnants would have internal regions
contracting from
internal past null or timelike infinities.) Then remnants could never
be
destroyed
(as anti-remnants could by the $CPT$ reverse of the process
of remnant formation), although presumably they could merge
or be swallowed by black holes (which could later become
remnants again).

     In this latter possibility, it is presumably true
\cite{HawHay,Haw93pri} that a fixed finite energy in a box would
never evolve into an absolutely $CPT$-invariant thermal state,
since there would eventually tend to be more remnants than
anti-remnants.  However, contrary to \cite{HawHay,Haw93pri},
I see nothing violating $CPT$-invariant evolution in this
scenario.  Furthermore, if from the outside remnants don't look
much different from anti-remnants (as would be the case for
cornucopia that tend toward static configurations as seen from
the outside), the state of the box at late
times could appear very nearly $CPT$ invariant, particularly
since one can readily estimate that it would be exponentially
rare to have more than one remnant in the box (assuming
they can merge or fall into a black hole).

     Possibility (3) might be subdivided into case (a) in which
information could never be retrieved from inside the remnants,
case (b) in which some, but not all, of the information could
in principle be retrieved, and case (c) in which all of the
information could in principle be retrieved.  In case (a) all
of the remnants presumably would be distinct from
anti-remnants, in (b) some, but not all, of the remnants would
be, and in (c) apparently there would be only a single remnant
state that would be distinct from its $CPT$-reversed
anti-remnant.  Cornucopia with internal future null or timelike
infinities, where information can go and never be retrieved,
would presumably fall into case (a), unless some of the internal
information does not go to the future null or timelike infinity,
in which case they might fall into case (b).

     (G) Evolution in which the disappearance of black holes results
in mixed
states that are unpredictable.  At the time, I did not have any
proposed model
for this, but later I realized \cite{Pag82,Pag83} it could occur for
a
$CPT$-invariant model in which our universe is an open system, and
information
can both leave and enter.  An analogue would be a room with a window:
from the
density matrix of the inside of the room alone at one time, one
cannot know
what light might come in from the outside, and hence one cannot
predict even
the density matrix inside the room at a later time.  Unlike the case
of
deterministic evolution of the density matrix by a generic
superscattering
matrix, in the case of an open system one generally cannot
extrapolate backward
from the later density matrix to a unique earlier one, so information
would be
truly lost in an even more fundamental way.

     (H) Replacement of density matrices by something more
fundamental.  I had
no proposals for this, but in the Letter I did say, ``In view of the
historical
developments in the concept of nature, one might say that the most
radical,
(H), is the most realistic.''  The interpretation of taking the
superscattering
matrix as merely being a tool to calculate conditional probabilities
would in
some sense require this, so perhaps Hawking's proposal of information
loss
would fit better here than under (B).  It is now also extremely
interesting to
see a whole new formal approach being developed, the decoherent
histories
reformulation of quantum mechanics, particularly the generalized
quantum
mechanics without states [74-78].  If it is indeed the
correct approach, it should have something to tell us about black
holes and
information.

     In the debate I thus initiated with Hawking, informal
discussions showed
me that relativists tended to side with Hawking's viewpoint, often
arguing that
the event horizon should be an absolute barrier to the recovery of
information,
whereas particle physicists were much more sympathetic to the
possibility of a
unitary S-matrix.  For example, Witten \cite{Wit81pri} dismissed the
idea of
the horizon as a barrier, since the uncertainty principle applied to
gravity
would prevent its localization.  However, in the early years there
were very
few papers about the subject.

     One interesting earlier paper that was never even published was
Dyson's
suggestion that if information were lost from our universe, it might
simply go
into what we would now call a baby universe rather than being
destroyed at a
singularity \cite{Dys76}.  (Around the same time, Zel'dovich
\cite{Zel77a}
suggested that baryons could leave our universe and go into a closed
space by
the Hawking process, but he did not discuss information there, and,
as noted
above, he later argued \cite{Zel77b} against Hawking's proposed loss
of information.)  I have already cited Wald's papers [52-54]
that argued that $CPT$ should indeed be broken in a strong sense and
pointed
out how it could be preserved in a weak sense by a superscattering
operator.

    A conference abstract of mine \cite{Pag80b} noted that another
problem
with
the semiclassical approximation is that if one feeds a black hole for
a
sufficiently long time at the Hawking emission rate, the SCA gives an
arbitrarily
large number of internal configurations for a black hole of a given
size, not
just
the exponential of the entropy Hawking had derived for it
\cite{Haw75,Haw76a}.
Furthermore, it seemed to me that even without $CPT$ invariance in
the strong
sense, it would be most natural to assume that the Hilbert spaces on
the
initial and final surfaces would have the same dimension, and then
(at least if
those dimensions were finite) the postulated existence of the 3-index
$S$
matrix of Eq. 6 would imply that the hidden hypersurface could only
have a
trivial Hilbert space (one unique state $A$).  Then the sum over $A$
in Eq. 6
would collapse to a single term, giving a unitary superscattering
matrix.

     The lack of $CPT$ invariance in the strong sense for a
superscattering
operator and the problem with the dimensions of the Hilbert spaces
motivated me
to consider the example of possibility (G), that our universe is an
open
system, with not one but two hidden Hilbert spaces, one from which
states can
come (say baby universes in the past), and one to which states can go
(say baby
universes in the future) \cite{Pag82,Pag83}. One could postulate that
there
is
an $S$ matrix, from the product Hilbert space of our past universe
and the past
baby universes, to the product Hilbert space of our future universe
and the
future baby universes.  If the initial density matrix on the past
product
Hilbert space were a tensor product of a fixed density matrix for the
past baby
universes and an arbitrary density matrix for our past universe (so
the baby
universes started uncorrelated with our universe), then the final
density
matrix of our universe would indeed be given by a superscattering
matrix
(depending on the initial baby universe density matrix, but that is
assumed
fixed) acting on the initial density matrix of our universe, which
would be
possibility (B).  This would be like the case of a room with a pure
state
outside, say the completely dark vacuum state, or perhaps a fixed
thermal state
which is completely uncorrelated with what is in the room.  In this
case, from
knowing the initial state of what is inside, one can give the density
matrix of
what will bounce back from or enter the window.  At the other
extreme, if the
final density matrix on the future product Hilbert space were a
tensor product
with a fixed density matrix for the future baby universes, then there
would be
evolution backward in time by a reversed superscattering matrix,
possibility
(C).  But in general, if the baby universes are correlated with our
universe in
both the past and the future, or if their statistical state is
unknown, one
would not have a superscattering matrix at all, possibility (G).

     This possibility in some sense sounds the most likely,
especially if
quantum gravity can allow connections to baby universes that can
branch off or
join on.  However, it raises some questions that are not yet very
clear.  For
example, if the dimension of the Hilbert space of our universe stays
the same
from past to future, then the two hidden Hilbert spaces should also
have the
same dimension in order that there be an $S$ matrix between the two
product
Hilbert spaces (at least the argument would be valid if all these
dimensions
were finite).  That would mean that there would be in principle as
many ways
for information to enter our universe as to leave it.  And yet the
semiclassical approximation seems to show many ways for old
information to
leave our universe (e.g., by going down a black hole), but the only
place it
seems to allow for new information to enter is at a possible naked
singularity
at the end of the black hole evaporation, where the semiclassical
approximation
breaks down.  One might even expect quantum gravity to heal the naked
singularity so that no new information enters the universe from it, a
possibility
I termed Quantum Cosmic Censorship \cite{Pag80}.  In other
words, the semiclassical approximation suggests that the dimension of
the
future hidden Hilbert space is large, but that the dimension of the
past hidden
Hilbert space is small or perhaps even zero.  If the two are actually
equal,
which suggestion is correct?  Taking the large dimension supports the
view that
pure states go to mixed states, but taking the small dimension
suggests that
little or no information is lost, and that pure states may stay pure.
This is
one of the strongest reasons for me to think that possibility (A),
unitary
evolution, is at least reasonably likely, and to resist what often
seem to me
to be premature reasons to dismiss it.

     On the other hand, it could turn out that even if the dimensions
of the
two hidden Hilbert spaces are identical and nontrivial, some
principle
influencing the states on those two spaces might make it so that in
actuality
more information leaves our universe than enters it.  This is
apparently
happening in my office room now at night as I type this, for outside
it is
dark, and little information in the visual band of photon modes is
coming in,
whereas there is much more information going out from the light
inside.  From
the inside, I can more easily predict the light I now see (reflected)
in the
window, whereas in the daytime, I cannot predict the light entering
from the
clouds outside I see floating past.  So in this language the question
would be,
why do past baby universes seem to be dark?

     Perhaps the answer is that something like
the Hartle-Hawking no-boundary proposal
[83-86],
the Vilenkin tunneling proposal
[87-94],
or the Linde inflationary proposal
[95-98]
makes the state of small past baby universes
simple, just as the state of our past universe seems to have been
simple when
it was small.  Now our universe has grown to be large and
complicated, and so
if it connects to the Hilbert space of small baby universes in
initially simple
states, information would naturally tend to go from our universe into
the baby
universes rather than the other way around.

     It would be interesting to try to formulate the problem of black
hole
information in terms of the no-boundary proposal, at least if one
could avoid
being stymied with all of the problems of the path integral for
gravity.
Assuming that the path integrals are over some contours of complex
geometries,
one would have to reformulate the concepts of ``initial'' and
``final,''
``past''
and ``future,'' etc. so that they are not in terms of the classical
Lorentzian
concept of time.  And then there is the question of whether it is
better to
treat the possible ``loss'' of information as a process to be seen in
a
decohering set of histories
[99-103, 74-78, 104],
or as something to be found in the records
existing in a single ``marvelous moment''
[105-122].
An interesting recent paper
by Smolin \cite{Smo} takes a quantum-cosmological approach to the
problem and
conjectures that if quantum effects do not eliminate singularities,
``loss of
information is a likely result because the physical operator algebra
that
corresponds to measurements made at late times must be incomplete.''

\vspace{5 mm}
{\bf 3.  Further Arguments for Four-Dimensional Black Holes}
\vspace{5 mm}

     After these rather few responses to Hawking's original proposal
for
information loss, some new interest was generated by a new paper
\cite{Haw82}
in which Hawking attempted to put the idea in an axiomatic framework
and apply
it to processes involving tiny virtual black holes.  He proposed a
set of
axioms for scattering in quantum gravity with asymptotically flat
boundary
conditions.  These included most of the usual axioms but omitted the
axiom of
asymptotic completeness, so that a pure initial state would not give
a unique
pure final state.  Alvarez-Gaum\'{e} and Gomez \cite{Alv83} gave a
rigorous
derivation of $CPT$ from Hawking's axioms and pointed out some
difficulties
with Hawking's idea of asymptotic incompleteness.  Gross \cite{Gro84}
showed
that nontrivial topologies in the path integral for quantum gravity
need not
give asymptotic incompleteness and a loss of information, though
Hawking
\cite{Haw84} argued that more complicated examples than the ones
Gross
considered would.  Ellis, Hagelin, Nanopoulos, and Srednicki
\cite{Ell84}
noted
that with a superscattering matrix rather than an $S$ matrix,
symmetries no
longer imply the usual conservation laws, so that the latter would
need to
added independently.  Banks, Peskin, and Susskind \cite{BPS} showed
that
making
the superscattering operator act locally would lead to a violation of
energy-momentum conservation.  Hawking's response \cite{Haw84} was
that it
should not be made into a local operator.  More recent studies
\cite{Sre,Liu}
have reaffirmed  and intensified various aspects of this problem but
do not
seem to have conclusively shown that no formulation with loss of
information
could be consistent with energy-momentum conservation.

     In addition to the proposals for loss of information, and the
ensuing
counterarguments, a program on the other side,
pursuing possibility (A) above, was launched by  't Hooft
[16, 21-23, 25-28, 31].
He made a bold analogy between string
worldsheets and event horizons and attempted to work out the
principles for an
$S$-matrix for black hole processes.  This has provided some
innovative ideas
of how information might be preserved, but it is unfortunately
probably too
optimistic to expect this program to be brought to completion in the
foreseeable future, because of the severe difficulties of quantum
gravity.

     Another development that appeared to tip the balance somewhat
toward
possibility (A) is the work on quantum wormholes and baby universes
[125-127, 11, 128-138, 112, 139-146].
This primarily addressed the question of the cosmological constant,
but it also
addressed the question of whether information can get lost into
wormholes.  The
somewhat surprising answer was that although there may be different
superselection sectors of the theory (each with different low-energy
effective
coupling constants that would have to be determined experimentally
rather than
theoretically from some fundamental `theory of everything'), in each
sector one
would get an $S$ matrix with no loss of information (though
Strominger has
recently argued \cite{Strom93}
that in each sector one would get an $\$$ matrix).  Few of the
experts claimed
that black hole formation and evaporation could be described by
wormholes, but
Hawking did \cite{w6}, which seemed to undermine his proposal for
loss of
information.  He did try to argue that the uncertainty of the
coupling
constants represented the loss of information \cite{w10,w15}.
However, it
would
really be different from his previous proposal, in that once one did
enough
repeated black hole collapse and evaporation experiments to measure
the
relevant effective coupling constants (assuming there were only a
finite number
that had any significance for the process at hand), one could predict
final
pure states for any subsequent experimants.  An interesting question
would be
the number of relevant effective coupling constants for a certain
process, and
only infinity would correspond to the continual uncertainty of
Hawking's
original proposal.

     To illustrate this difference, consider a simplified process in
which
there is a finite box containing either a ``red'' or a ``green''
particle of
the
same mass, each of which would be absolutely stable in the absence of
gravity.
With gravity, suppose each could collapse to form a black hole which
could then
emit either kind of particle.  (If you do not believe one particle
could do
this, replace it by two particles or whatever you think the minimum
number is.)
For simplicity, assume that the black hole can emit its energy in no
other
forms.
In Hawking's original proposal, an initial  pure state of red, say,
would
become mixed and would tend toward the mixture of 50\% probability
for red (but
never lower) and 50\% for green (but never higher).  However, in the
wormhole
analysis, there would presumably be an effective coupling constant
for the
transition rate from red to green and vice versa, leading to a
coherent
oscillation between red and green in each superselection sector.
{\it A
priori}, one might not know this rate and therefore not be able to
predict
better than a density matrix for the final state.  However, after
sufficient
measurements of the rate to find out which superselection sector one
is in, one
could predict thereafter the oscillation rate.  In particular, one
could
predict times at which the originally red particle has a probability
of unity
to be green (to the accuracy of the measured coupling constant), a
situation
that would never occur for Hawking's superscattering matrix.

     Here I have been assuming that there is only one relevant
coupling
constant for the transition rate.  Of course, there might be more,
say
depending on the state of the particle in the box.  If there were
were an
infinite number of relevant coupling constants, and if they coupled
to states
outside the box (as they presumably would have to if there are only a
finite
number of relevant possible states within the box), then one might
never be
able to predict a probability greater than 50\% for a green particle.
This
might
happen, for example, if the coupling were sufficiently strong to the
records of
the previous measurements so that they become anti-self-fullfilling
prophecies:
 the results of repeated experiments would be identical only if there
were no
records to mess up the subsequent experiments, but if no records were
kept,
then no one could predict the repetition.

     At one time I thought Hawking's suggestion of describing black
hole
formation and evaporation by a wormhole process was reasonable, and
then it
seemed that there would be no loss of information once enough
measurements
were made (with various speculations of how many might be needed)
\cite{Pag92}.
 However, it now seems to me that the standard wormhole calculus is
probably
not applicable \cite{Pag93}, since it involves integrating over
length scales
up to the wormhole length in order to get an effective theory at
larger scales.
With black hole formation and evaporation that would take a timescale
of order $M^3$ in Planck units (assuming no long-lived remnant at the
end,
which would only intensify the problem), it
would seem that one would need to integrate over length scales up to
about
$M^3$, and the effective theory would apply only on length scales
larger than
that.  But then the effective theory could not describe what
collapsed to form
the hole (in a much shorter time) or the individual quanta being
emitted (with
wavelength of order $M$).

     Bekenstein has recently conjectured \cite{Bek93} that since
Hawking radiation is not precisely blackbody but rather is greybody
(because
of the partially reflecting curvature and angular momentum barriers
around
a black hole), this nonthermal aspect could code the
information in the black hole.  However, it does not seem that this
effect,
or another similar nonthermal effect (such as stimulated emission
when
incident radiation is present \cite{Schif93}),
could by itself ever lead to a pure final state \cite{DanSch,MulLou},
since each of them occurs already in the semiclassical
or even classical approximation.  Indeed, the
generalized second law for quasistationary black holes
[152-155, 1, 2, 156-184]
implies that under these approximations the entropy of the radiation
would
always be at least as large as one quarter the area lost by the black
hole.
The information actually in the radiation simply means that it is not
completely
random.  For example, it might be such that from the complete final
state,
even if it is an impure density matrix, one could in principle deduce
the
complete initial (possibly pure) state, as was discussed above.
However,
this does not exclude the possibility that the information in the
sense of Eq.
(2) above might decrease.

     Another attempt to avoid a loss of information in black holes is
to
postulate that black holes never really form.  For example, Frolov
and
Vilkovisky \cite{FV79,FV81} and H\'{a}j\'{\i}\u{c}ek
\cite{Haj86,Haj87}
conjectured that gravitational collapse might lead to no
singularities or
event horizons (only apparent horizons), and so no true black holes.
Nevertheless, there would be a very large time delay before ingoing
null rays become outgoing null rays, and there would be Hawking
radiation,
so the quantum-corrected system would appear much like a
true semiclassical black hole, thus fulfilling the correspondence
principle.
Unfortunately, our understanding of quantum gravity is too meagre
at present to confirm or refute this conjecture, at least in four
dimensions.

     An even more direct way to try to eliminate black holes is to
assume
a different classical theory of gravity.  For example, Moffat
\cite{Mof93,Mof93b} has postulated that if the correct theory of
gravity
were NGT rather than GRT, the NGT charge could prevent black holes
from forming.  But even if NGT were
a consistent theory of gravity [189-193],
it would allow black holes to be formed from pure radiation without
NGT
charge, and so it would not really succeed in circumventing the
problem.
It would probably be very difficult for any simple consistent
classical theory
of gravity (which agrees with Newtonian gravity and with special
relativity
in the appropriate limits) to avoid producing black holes in all
circumstances.

     Other arguments have been given [194-197] that
black holes have their area $A$ quantized in units of $CL_{Pl}^2$,
where $L_{Pl}$ is the Planck length (set equal to unity), and $C$ is
a numerical constant of order unity (e.g., $8\pi$ \cite{Bek74b},
$4\ln 2$ \cite{Muk,GarBel},
or $16/(3\pi)$ \cite{Peleg}.  This quantization of the black hole
would seem
to make
most sense if there were unitary evolution, with the black hole as
part of the
intermediate quantum state, and with no loss of information.
However, these
proposals appear to imply \cite{Muk} that an uncharged black hole
with zero
angular momentum and mass $M\gg 1$ (which has an area $A=16\pi M^2$,
at least classically) would have the nearest different energy level
differing
by
an energy very nearly $C/(32\pi M)$.  Then the black hole could
absorb or emit
single quanta of radiation with energies only integer multiples of
this
(i.e., wavelengths equal to $64\pi^2 M/C$, which is of the order of
the
Schwarzschild radius $2M$ of the hole, divided by an integer, for
radiation
quanta of zero rest mass).  This line spectrum \cite{Muk} would be
significantly different from the
expected semiclassical limit in which one gets quantum field theory
in the
classical spacetime of the black hole, which would allow absorption
or
emission of a continua of frequencies or wavelengths (e.g.,
wavelengths
longer than $64\pi^2 M/C$).

     A loophole in this argument is the possibility that the
classical relation
between area and mass is invalid, so that an area quantization does
not
imply the na\"{\i}vely corresponding mass quantization.  But in any
case,
if the black hole mass is quantized,
I would expect the levels generally to be discrete and be separated,
on
average, by the inverse of the level density (e.g., by roughly
$e^{-A/4}$,
or $e^{-4\pi M^2}$ for uncharged black holes with zero angular
momentum),
rather than being clumped in highly degenerate levels that are much
further
separated (e.g., by $C/(32\pi M)$).  (Actually, I would expect the
levels
to be discrete only in the case that one put the black hole in a
finite box and
considered the energy levels of the entire system in the box, e.g.,
the
black hole plus the surrounding radiation.  For a black hole in
infinite space,
the instability to evaporation would smear out the levels by amounts
much greater than their separations if the separations were indeed of
order
$e^{-A/4}$.)  If my expectation were true, it would seem quite
possible to
get, in the semiclassical limit, ordinary quantum field theory in the
curved
spacetime of a classical black hole, with no noticeable line spectra
or
departures from the expected ordinary thermal spectra.

     There has also been a long sequence of analyses of black hole
evaporation
in string theory, and related processes, by Ellis, Mavromatos, and
Nanopoulos
[198-210].
The latest conclusion seems to be \cite{EMN13} that there is an $\$$
matrix
with loss of information.  However, I am not competent to judge this
work,
and there doesn't seem to be much other comment on it in the
literature that
I am aware of.

\vspace{5 mm}
{\bf 4.  Two-Dimensional Black Hole Models}
\vspace{5 mm}

     After nearly sixteen years in which only a few papers per year
were
written directly about Hawking's proposal, the subject experienced a
strong
upsurge of interest as a result of a paper by Callen, Giddings,
Harvey, and
Strominger \cite{CGHS}.  This paper replaced the presently
intractable problem
of four-dimensional gravity with a two-dimensional model theory,
motivated by
extreme dilatonic black holes \cite{Gib88,GHS} and by string black
holes
[214-218].  This two-dimensional model is much simpler
and yet seems to capture many of the aspects of four-dimensional
gravitational
collapse and evaporation.  One might object that this toy model might
miss the heart of the problem in four dimensions, but since we do not
yet have
any consistent understandable theory of quantum gravity in four
dimensions, any
attempt to understand the problem must make some simplification or
truncation
of the unknown complete theory, and so one might as well start first
with the
simplest such model that apparently has enough of the realistic
features.
Although the two-dimensional model  does not really have independent
gravitational degrees of freedom, it does have black holes and matter
degrees
of freedom (the dilaton and minimally coupled scalars), which can
give Hawking
radiation and lead to at least some measure of black hole evaporation
when the
back reaction of this radiation is included.

     This simplified model can be solved exactly classically.
However, its
extension to a consistent quantum theory appears to have an infinite
number of arbitrary parameters
[219-236, 30, 237-247, 46].
In some cases the quantum theory
may be be solved exactly, and in a wider class of cases the
semiclassical
equations are analytically soluble, but no completely satisfactory
quantum
model has been yet found.  Much of the work has concentrated on
solving
the semiclassical equations of the original CGHS \cite{CGHS} model.
At first it was thought \cite{CGHS} that there would be nonsingular
evolution with an $S$ matrix, but then it was shown
[62, 220, 248-254]
that the semiclassical equations generally lead to singularities.
It is not yet known what this
model, or the various modifications of it that have been proposed,
would really
give in a full quantum analysis concerning information loss.
However, this
model has inspired much new thought about the subject, some of which
I shall
now summarize.

     One approach, which is somewhat motivated by and patterned after
't Hooft's program
[16, 21-23, 25-28, 31]
in four dimensions, is to look for a unitary
$S$-matrix for two-dimensional black hole formation and evaporation
\cite{Mik,Ver93,Sch93,Rus93,Mik93}.  This can apparently be done,
for example \cite{Ver93,Sch93}, by imposing reflecting
boundary conditions at a critical value of the dilaton field, though
there are
some subtleties (e.g., when this critical value occurs on a spacelike
line),
and of course there is the question of whether these boundary
conditions are
realistic.  However, it is interesting that this work seems to give
an explicit
example of a process that looks very much like black hole formation,
followed
by nearly thermal Hawking radiation, and yet is described by an
$S$-matrix with
no loss of information, possibility (A), that I have continually
argued has
remained a viable alternative.  Unfortunately, it appears that the
full $S$-matrix of \cite{Ver93,Sch93} may not be unitary, and the
part that
is may merely represent the part in which no black hole forms
\cite{Suspri}.

     A second series of papers patterned after 't Hooft's program is
\cite{STU,Sus,SusT},
which proposes that there is a ``stretched horizon,'' a membrane just
outside the global event horizon which appears to be physically real
to an outside observer.  From the outside viewpoint,
``{\it the stretched horizon is a boundary surface
equipped with microphysical degrees of freedom that appear in the
quantum Hamiltonian used to describe the observable world}.''
As a quantum object, the stretched horizon may be able to store and
return all of the information of the gravitational collapse to the
outside.

     It is not clear in this approach how the information gets from
the
infalling fields to the stretched horizon.  It is not proposed that
an
infalling observer would feel his or her information getting
bleached,
so that it would all immediately go onto the stretched horizon
as he or she crosses it.
Instead, the claim is that ``there is {\it complementarity\/}
between observations made by infalling observers who cross the
event horizon and those made by distant observers.''  Preskill's
view of this \cite{Prepri,Suspri} is that any apparent contradiction
which arises when one tries to combine the viewpoints of infalling
and outside observers will actually involve assumptions about
physics at energies above the Planck scale \cite{SusT}.
Nevertheless, it would be useful to have a mechanism for seeing
how in detail the information can get onto the stretched horizon.

     Another approach, which lends some support for the opposite
conclusion, is
the study of scattering by extremal black holes
[257, 211, 258, 62, 259-261, 63, 64, 262, 263, 65].
It does seem from these analyses that if one had a theory
with absolutely stable large extremal black holes
(e.g., electrically charged holes in a theory with no charged
particles,
such as Einstein-Maxwell theory with only neutral other fields, or
magnetically
charged holes in a theory with no magnetic monopoles), then there may
well be
an infinite number of arbitrarily low-energy perturbation states deep
down in
the throat of these black holes.  One could presumably lose an
arbitrary amount
of information into these states, though it would be a bit ambiguous
whether
one said this information were lost from our universe or persisting
in the
remnant.

     To be relevant to the problem of black hole formation and
evaporation, one
would have to assume that these stable black holes are not merely
eternal parts
of a background spacetime but can be produced by some process.  If
the black
holes are charged in a theory with no charged matter, one could not
form them
from the gravitational collapse of matter.  One might instead imagine
forming
them by pair production.  This raises the problem that if there are
an
infinitely large number of these remnant states below a fixed finite
energy,
then it would appear that they should be produced at an infinite rate
\cite{ACN} (e.g., in the Hawking radiation of a larger black hole).
This has
appeared to be a very strong argument against an arbitrary amount of
information from gravitational collapse going into a remnant of
bounded mass
(e.g., the Planck mass).

     However, possible reasons have been given why the standard
argument might
be wrong and that an infinite degeneracy of remnants can be produced
at a
merely finite rate.  The first reason given \cite{ACN} was that the
remnant
form factors might vanish when the momentum transfer is timelike.
This
possibility, which I am not competent to judge, seems to have been
generally
ignored in the literature.  Another reason given is that the remnants
may have
infinite internal volumes which can carry an arbitrarily large amount
of
information while being pair produced at a finite rate
\cite{BO'L,BO'LS,ST,LO'L}.
On the other hand, there are counterarguments \cite{Gid93} that
unless there
is
what would be described from the effective remnant
theory as a ``strong coupling conspiracy,'' there would still be an
infinite
production rate.  The possibility of an arbitrary amount of
information
remaining in remnants seems to be more open than I would have thought
it was
two years ago, but it is yet by no means convincing to me.

     Another remnant scenario is Giddings' proposal \cite{Gid92} that
the
information from gravitational collapse ends up in a remnant whose
mass and
size depend on its information content.  This would avoid the problem
of
infinite production, since there would only be a finite number of
states up to
any given mass (in a finite volume), and so any process with a finite
amount of
energy available could only access a finite number of states.
However, even
Giddings recognizes that these remnants ``would seem to require new
physics at
weak curvatures and runs afoul of causality'' \cite{Gid93}.

     Bekenstein \cite{Bek93b} has raised a related objection to
information-bearing remnants based on his conjectured limit
\cite{Bek81}
on the entropy $S$ of a system of bounded linear size $R$ and energy
$E$,
$S\le 2\pi RE$.  (This conjecture is supported by a number of
examples in flat spacetime
[158, 162, 165, 264-266, 173]
if one chooses judicious definitions for $S$, $R$, and $E$, and
it has somewhat weaker support
[159, 175-177, 181]
for self-gravitating systems, but it by no means seems to be proved
in general \cite{UW82,Unw82,Pag82b,UW83}.)
If the thermodynamic entropy $S$ of a remnant is indeed limited,
but if the information capacity (the number of possible internal
states)
is not limited,
one would have the same problem \cite{Pag80b} discussed above
for the semiclassical approximation applied to black holes of finite
thermodynamic entropy that are fed radiation
for a sufficiently long time at the Hawking emission rate.
However, if those who believe an arbitrarily large amount of
information can be put into a black hole in this way without an
equal amount coming back out have an answer for this
mystery (why the thermodynamic entropy remains bounded
while the information content of the hole grows indefinitely),
then the same answer would presumably allow remnants to
contain an arbitrarily large amount of information, even if their
thermodynamic entropy is bounded, say by Bekenstein's
conjectured limit.

     Even if large information-storing remnants can be shown to be
possible and
truly consistent in some model theory, I am sceptical that they would
occur in
a realistic theory of gravity.  Extremal electrically charged black
holes in
the real universe are unstable to emitting charged particles, such as
electrons
and positrons, by essentially the Schwinger process \cite{Sch51}
(assuming
that
such black holes are sufficiently large that a semiclassical analysis
suffices
for the emission rate) [269-271].  Extremal magnetically
charged
black holes would presumably be unstable by the analogous process
\cite{Aff81}
to emitting magnetic monopoles, which in typical grand unified
theories exist
with magnetic charge greater than mass and therefore could
energetically be
emitted by an extremal hole.  Once the magnetically charged hole
emits enough
of them to get small enough, there is even a classical instability
for it to
convert into a monopole-like configuration with a small black hole at
the
center which then gets hotter and hotter as it emits
\cite{LNW1,LNW2}. Only
if
no such magnetic monopoles existed would extremal magnetically
charged black
holes be absolutely stable.  I suppose that it is conceivable that
the true
theory of the world would allow some sort of extremal black holes but
no
suitably charged matter that could be emitted from them.  However,
until there
is more evidence for such hopes, it would seem more natural to
investigate theories in which black holes can decay at least down to
the Planck
mass.

     Since we do not know how to analyze what might happen then, one
could
postulate that the information remains in a Planck size remnant (if
they
somehow can avoid the infinite-production disaster), but we can as
yet do no
calculation supporting this hypothesis.  In fairness, one could say
this is no
worse than the hypotheses that black holes evaporate away completely
and that
information either is or is not lost, neither of which can be
presently
supported by calculations in domains where they can be believed.
Nevertheless,
in the absence of a clear mechanism for keeping Planck mass remnants
absolutely
stable, it would seem more natural to suppose that they would simply
decay
away.  After all, particles that are believed
to be absolutely stable all have some symmetry principle preventing
their decay
(such as charge conservation for the stability of the lightest
charged
particles, electrons and positrons).  Unless there is some unknown
symmetry
principle protecting a Planck mass remnant, it would seem highly
surprising for
it not to decay.  Indeed, Zel'dovich \cite{Zel77a} and Hawking
\cite{Haw82}
argued that remnants of primordial black holes would unacceptably
dominate
the mass density of the universe. However, I am not aware that this
argument
was ever presented in the days before black hole evaporation was
discovered,
when it would have presumably have been an even more serious problem.
Furthermore, inflation would most likely make the density of the
remnants
of primordial black holes quite acceptable even if they are massive.
Still, if I had to bet, I would lay low odds on the possibility that
the
information
contained in gravitational collapse goes into stable massive
remnants.

\vspace{5 mm}
{\bf 5.  Can the Information Come Out in the Hawking Radiation?}
\vspace{5 mm}

     If one examines the alternative hypothesis, that all black holes
eventually evaporate completely (assuming that the universe lasts
long enough
and that it is not a $k=0$ universe in which black hole coalescence
always
continues to dominate, on average, over Hawking evaporation
\cite{PM1,PM2}),
then  we still have the uncertainty of whether the information is
lost or not.
Furthermore, if the information is not lost, there is the question of
whether
it comes out throughout the Hawking emission or whether it comes out
only after
the black hole gets down to near the Planck mass.  Hawking argued
\cite{Haw76}
that it would be impossible for the information to come out then, for
there
would not be enough energy, but that is only valid under what may
have been the
implicit assumption that the remaining energy comes out in a short
time.
Aharonov, Casher, and Nussinov \cite{ACN} proposed that the remnant
would
decay
in an exponentially long time, as Hawking also did later
\cite{Haw82}. Carlitz
and Willey \cite{CW87}, and later Preskill by a more general argument
\cite{Pre92}, showed that the actual lower bound for the lifetime of
a Planck
mass remnant, which contained all the information originally in a
black hole of
mass $M$, would be of order $M^4$ in Planck units, which is much
longer than
the period of order $M^3$ for the black hole to get down to near the
Planck
mass by Hawking radiation, but not exponentially longer.

     This slow decay of a remnant which contained all of the original
information seems to me no more reasonable than the absolute
stability of the
remnant.  It is hard for me to see why a Planck mass object could
last so long.  I can see that it would be required if the remnant
indeed had a
huge amount of information which has to escape as the remnant decays,
so what
actually seems unlikely to me is the possibility that a small remnant
can
contain a huge amount of information that eventually comes back out.
If it
could rigorously be shown that the information in gravitational
collapse does
not almost entirely come out before the black hole shrinks to a small
size, I
would think it more reasonable to suppose that the remaining small
object
disappears in a short time, without the emission of the information,
than for
the small object to spit out all the remaining information in the
long time
required.

     Therefore, it seems most probable to me that the black hole
decays away
completely, and that either the information comes out slowly during
the entire
emission process, or else it does not come out at all.  Various
problems have
been noted recently with the suggestion that the information comes
out, some of
which sound serious, so I shall now examine them.  It turns out that
none of
them appear airtight to me.

     Giddings \cite{Gid93} argues that this version of possibility
(A) ``would
require that the Hawking radiation extracts {\it all} information
from the
ingoing matter, {\it e.g.} through scattering as they cross near the
horizon.
In particular, matter that crosses the horizon and falls towards $r =
0$ must
therefore have essentially zero information content.  For this reason
this
option seems...farfetched to many.''  If there really is an absolute
global
event horizon, his argument seems valid, but I think it more likely
that the
quantum uncertainty applied to the causal structure of the spacetime
makes it
impossible to define exactly an absolute horizon.  The information
might be
taken out of the matter near what is interpreted classically as $r =
0$ and yet
not be, with 100\% quantum probability, within any putative absolute
horizon.
Or, the Principle of Black Hole Complementarity \cite{STU,Sus,SusT}
may somehow allow the information to fall towards $r = 0$ as seen by
a freely infalling abserver and yet appear to an outside observer to
stay
outside the horizon.

     Harvey and Strominger \cite{HS} allow for the fact that a global
event
horizon may never form but say, ``The big difference for black holes
(as
stressed in \cite{Wal84}) is that until the final Planckian stage of
the
evaporation they are surrounded by an apparent horizon which is very
nearly
null.  The infalling particles therefore carry the information into a
region
causally shielded from that part of future null infinity which
precedes (in
retarded time) the final stage of evaporation.  Thus the information
cannot
come back out without violating macroscopic causality until the black
hole has
evaporated down to the Planck size.  It is conceivable that quantum
coherence
could be restored by radiation emitted in the final stage of
evaporation which
is governed by unknown laws of quantum gravity.  However, since the
total
available energy is bounded and small (relative to the initial black
hole
mass), this is possible only is the radiation is emitted over an
extremely long
period \cite{ACN}.''  Although this argument allows for back reaction
so that
the apparent horizon is not truly an event horizon, it still takes an
essentially semiclassical view with a definite classical metric
(whose form
presumably depends on some quantum average of the Hawking emission).
True
quantum fluctuations in the geometry, and hence in the resulting
causal
structure, could invalidate the argument and allow information to
come out long
before the apparent black hole has shrunk to near the Planck size.

     Preskill \cite{Pre92}, building on discussions with Susskind and
earlier
work by Susskind and Thorlacius \cite{SusTho}, gives a longer
argument that I
shall only briefly summarize here.  He notes, ``On the spactime of an
evaporating black hole, it is possible to draw a single spacelike
slice that
crosses most of the outgoing Hawking radiation, and {\it also}
crosses the
collapsing body, well inside the (apparent) horizon.''  If the
Hawking
radiation
on the part of the slice outside the horizon is to be a pure state
for each
possible pure initial collapse state, the state of the body on the
part of the
slice inside the horizon must be a {\it unique} state (or else for a
generic
initial collapse state, one would get correlations between the
radiation and
the body, so neither would be in a pure state).  (This argument,
a more detailed version of what I had pointed out in \cite{Pag80b},
as discussed above, was apparently first given in this particular
form by
Susskind \cite{SusAsp}, is also discussed in further detail in
\cite{DanSch,STU}, and is related to the fact that ``a single
quantum cannot be cloned'' \cite{WooZur}.)

    To get this unique state, Preskill continues \cite{Pre92}, ``a
mysterious
force must bleach'' the information out of the body before it crosses
the
horizon, which ``is hard to imagine any reasonable way to achieve.''
Again, a
conceivable way out is the quantum nature of the causal structure.
The slice
considered here must be nearly null (in the semiclassical picture) to
cross
most of the Hawking radiation, and it may be impossible in the
quantum gravity
picture to say that it is definitely spacelike.  If it is not, one
would not
expect to have a tensor product structure of Hilbert spaces for the
radiation
outside and the body inside, on which the detailed argument depends.
(A
slightly different-appearing  way out is the Principle of Black Hole
Complementarity \cite{STU,Sus,SusT} discussed briefly above.)

     Nevertheless, if the information almost all comes out
before the apparent black hole gets down near the Planck mass (which
I
am arguing is still a viable possibility), then I would
agree with Preskill's final caveat about this possibility:  ``At the
very
least, the semiclassical picture of the causal structure must be
highly
misleading.''

     Giddings and Nelson \cite{GidNel}, and later Giddings alone
\cite{Gid92b},
repeated some of the other arguments against the possibility that the
information comes out gradually throughout the emission (``Objection:
this
would appear to imply that either all of the information has been
extracted
from the infalling matter by the time it crosses the horizon, or that
information propagates acausally from behind the horizon to outside''
\cite{Gid92b}) that I have already countered here, but they also gave
a more
detailed version of one of Hawking's original arguments \cite{Haw76}.
Starting
from a semiclassical calculation for two-dimensional black holes in
which
information apparently would be lost, they noted that this analysis
``may, of
course, be invalidated once higher-order quantum corrections are
taken into
account.  However, these corrections are expected to be unimportant
until the
weak-coupling approximation breaks down.  This only happens in the
final stages
on the black-hole evaporation.  The above arguments therefore
strongly suggest
that within the present model information does not escape until the
black hole
is very small.  Making these rigorous will therefore rule out one
suggested
resolution of the black-hole information problem, namely, that the
information
escapes over the course of black-hole evaporation if the effects of
the back
reaction are included'' \cite{GidNel}.  Later Giddings toned this
down to a
more
tentative claim, that ``working order-by-order in $1/N$, it is
probable that
one can construct an argument...analogous to stating that the
information
doesn't come out of four-dimensional black holes until they reach the
Planck
scale'' \cite{Gid92b}.

     However, it seems to me more likely that if the information does
indeed
come out gradually over the entire emission process, initially the
the rate of
information outflow may be so low that it would not show up in
order-by-order
(perturbative) analysis.  The two-dimensional moving mirror model
\cite{DavFul} analyzed by Carlitz and Willey \cite{CW2}
and by Wilczek \cite{Wil} show that it is
theoretically possible to have the early Hawking radiation {\it
exactly}
thermal, in a maximally mixed state with no information, but then
entirely
correlated with the late radiation so that the total state is pure.
(The
Carlitz-Willey model had the late radiation coming out only after the
black
hole had shrunk to a small remnant, but one could have the late
radiation be
simply the second half of the Hawking radiation at the usual rate, so
that the
information almost entirely comes out in the correlations before the
black hole
gets extremely small.)  An order-by-order analysis of the information
in the
early radiation would show no information getting out, but the
conclusion that
the information cannot get out until the black hole gets small would
be
invalidated by the nonanalytic change in the information rate at the
beginning
of the correlated late radiation.

     One might object that the case of exactly thermal local
radiation, with
correlations only between the first half and the second half, is an
extreme
case that is not at all plausible.  Therefore, I did a calculation
\cite{Paginfo} of the
correlations one might expect in the most crude approximation that
the state of
the total radiation from the black hole (once it has completely
evaporated) is
a random pure state consistent with the macroscopic expectations
(conservation
of energy, momentum, and angular momentum, in modes that appear to
come from
where the black hole was, etc.).  To calculate the information in an
early part
of the radiation, I assumed that this radiation and the black hole at
that
stage were subsystems making up a combined system in a pure state.
(This is reminiscent of the entropy inside an imaginary sphere
for a quantum field in the vacuum state \cite{Bom,Sred93}.)  I took
the
exponential of the coarse-grained black hole entropy ($s=A/4$, given
by
Hawking's semiclassical calculation \cite{Haw75,Haw76a},
using a lower-case $s$ to
denote a coarse-grained entropy or logarithm of the dimension of the
Hilbert
space of states) as an estimate of the dimension of the Hilbert space
of black
hole states of the same macroscopic characteristics, and the
exponential of the
initial coarse-grained black hole entropy ($s_0=A_0/4$) as an
estimate of the
dimension of the total accessible Hilbert space of the combined
system.  Then I
calculated the average fine-grained entropy of entanglement,
$S\equiv-Tr(\rho\ln\rho)$, of the
radiation subsystem, averaged over all possible pure states of the
combined
system.  The difference between this average entropy and the maximum
entropy
possible for this subsystem, as given by Eq. 1 above, represents the
average
information of the subsystem.

     If the two subsystems have Hilbert space dimensions $m$ and $n$,
with
$m\leq n$, then, based on previous work by \cite{Lub,Llo}, I found
\cite{PagEnt}
evidence for a conjecture that the average entropy of each subsystem
is
     	\begin{equation}
	S_{m,n}=\sum_{k=n+1}^{mn}\frac{1}{k} - \frac{m-1}{2n}.
	\end{equation}
(Note added after publication:  This conjectured formula was shortly
thereafter proved by Foong and Kanno \cite{FooKan}.)
For $1\ll m\leq n$, I found (by reasoning independent of, but
consistent with,
the conjecture) that the average information in the smaller subsystem
is
	\begin{equation}
	I_{m,n} \simeq \frac{m}{2n}.
	\end{equation}
This means that the smaller subsystem typically has very little
information in
it.  That
is, for a typical pure quantum state of a large system, the smaller
subsystem
is very nearly maximally mixed, showing little signs that the total
system is
pure.

	Another way of summarizing this result is to say that if the
two subsystems,
with large Hilbert space dimensions $m$ and $n$, are
broken up into tiny sub-subsystems, which typically would each be
very nearly
maximally mixed, there would be virtually no information is the
sub-subsystems
considered separately.  For quantum information, the whole system
contains more
information than the sum of the information in the separate parts,
and in this
case almost all the information giving the precise pure state of the
entire
system, $\ln m+\ln n$, is in the correlations of the sub-subsystems.
The
result above shows that for a typical pure state of the entire
system, very
little of the information, roughly $m/(2n)$, is in the correlations
within
the smaller subsystem itself, roughly $\ln n -\ln m +m/(2n)$ is in
the
correlations within the larger subsystem itself, and the remaining
amount of
information, roughly
$2\ln m -m/n$, is in the correlations beween the larger and
smaller subsystems.

	Suppose we apply this to the case of a black hole which has
lost less than
half its initial coarse-grained entropy $s_0=A_0/4$ to radiation, so
	\begin{equation}
	m\simeq e^{(s_0-s)}\leq n\simeq e^s.
	\end{equation}
Then, using the average above as a typical value, a typical value for
the
information in this early part of the radiation would be
	\begin{equation}
	I_{rad} \simeq \frac{1}{2}e^{-(2s-s_0)},
	\end{equation}
which is exponentially small.  For example, if a Schwarzschild black
hole
decays primarily by emitting massless Hawking radiation so that its
mass
decreases at the rate
	\begin{equation}
	-\frac{dM}{dt}=\frac{\alpha}{M^2},
	\end{equation}
then the initial rate of information outflow from the hole (when
$M=M_0$) is
	\begin{equation}
	\frac{dI_{rad}}{dt} \simeq 4\pi\alpha M^{-1}e^{-4\pi
M^2}\simeq 4\pi\alpha x
e^{-4\pi/x^2},
	\end{equation}
where $x=M_{Planck}/M$ is a natural expansion parameter of how good
the
semiclassical approximation is for large four-dimensional black
holes.

	For two-dimensional dilatonic black holes, when the quantum
corrections
are small so that the classical equations of \cite{CGHS} provide good
approximations for the various thermodynamical quantities, the
Hawking
temperature is $T = \lambda/2\pi$ (independent of the mass $M$), so
the
semiclassical coarse-grained entropy is $s=2\pi M/\lambda=2\pi
e^{-2\phi_H}$,
where $\phi_H$ is the value of the dilaton at the horizon.  With $N$
minimally
coupled scalar fields, the quantum-corrected equations \cite{CGHS}
are valid
outside the horizon for $e^{-2\phi_H}\geq N/24$ or $s\geq \pi N/12$,
so the
minimum coarse-grained entropy of the black hole is bounded below by
a
constant times the number $N$ of scalar fields.

	Therefore, the information in the radiation when its
coarse-grained entropy
is $s_{rad} < s$ is
	\begin{equation}
	I_{rad}=s_{rad}-S_{rad} \simeq
\frac{1}{2}e^{s_{rad}-s}<e^{s_{rad}-\pi N/12}
	=e^{s_{rad}-1/x},
	\end{equation}
where now the small parameter is $x=12/(\pi N)$.

	Thus we see that the initial rate of information outflow is
not analytic in
the small expansion parameter $x$ at $x=0$, so one would never find
it by the
order-by-order (perturbative) analysis that Giddings and Nelson
advocate.
That is, even if they succeed in their goal of proving that the
information
does not
come out at any finite order of the perturbation, it would not be a
convincing
argument that the information is not actually coming out in a
nonperturbative
way, since that seems to be the typical behavior for a random joint
pure state
of a black hole plus radiation.

	Even if one looked at most of the radiation so that $I_{rad}$
is not
exponentially small in $1/x$, one would not see the information
unless one
made at least of order $e^s$ measurements, i.e., more than of order
$e^N$
measurements in the case of radiation from a two-dimensional
dilatonic black
hole.  The $1/N$ expansion may indeed still be good until $s$ gets
down
of order $N$ \cite{Gidpri}, so that one can in principle calculate
each
typical
measurement to an accuracy that is an arbitrarily high power of
$1/N$.
However, predicting the more than $e^N$ measurements necessary would
presumably be impossible to enough accuracy by this perturbative
method.
I.e., suppose that the rms error of each measurement could be made
smaller than any finite power of 1/N.  But when one squares and sums
the errors for more than $e^N$ measurements, one does not have
a result that can be controlled by making $N$ large in this
perturbative
analysis.  Therefore, this perturbative analysis simply could not say
whether the information is there or not.

	Of course, it must be admitted that although there does not
seem to be
any way at present to rule out the possibility that the information
in
gravitational
collapse comes out slowly during the entire period of radiation, no
really
plausible mechanisms have been proposed for this either.
The proposals of 't Hooft
[16, 21-23, 25-28, 31],
Mikovi\u{c} \cite{Mik,Mik93}, the Verlindes \cite{Ver93}, Schouten
and the
Verlindes \cite{Sch93}, and Russo \cite{Rus93} are possible
mechanisms,
and that of \cite{STU,Sus,SusT} is a framework without yet a
mechanism,
but it is by no means certain that any of these would work in
realistic
situations.

	    Since we do not yet have a definitive mechanism for
getting the
information out from a black hole, permit me to give a few of my own
speculations.  One idea would be that the information comes out in
wormholes
from the high curvature region (i.e., near $r=0$) to just outside the
(apparent) horizon.
(Such a suggestion of wormholes with one end inside a black hole has
been made by Frolov and Novikov \cite{FroNov}, though they
discuss the case of a small number of macroscopic wormholes,
whereas I wish to consider the possibility of a huge number of
microscopic quantum wormholes.)
One must of course ask why the information-transmitting
wormholes would tend to have ends at those locations, since by the
wormhole
calculus, they are supposed to have equal amplitudes of attaching on
anywhere
[135-137, 112, 139, 141].
However, the amplitudes can be affected by the {\it conditions} at
the
locations, and one could easily imagine that the high curvature could
increase
the amplitude for wormhole ends to be there.  At the other putative
ends, near
the horizon, there do not appear to be any special local conditions
to increase
the amplitude for having wormhole ends there.  However, one can have
pairs of
particles coming out of wormhole ends there without violating the
conservation
of energy, since one particle can escape to infinity with positive
energy,
while another carries a corresponding amount of negative energy down
the hole.
That is, in the path integral the integration over time that leads to
the
conservation of energy cancels the effects of all wormhole ends that
emit
but do not absorb particles far outside any horizon (or ergosphere)
but leaves the effect of such wormhole ends inside and at
the surface of the black hole.  Wormhole ends entirely inside the
apparent
black hole are ineffective in bringing information to the outside, so
only
wormhole ends near the horizon are unabated by energy conservation
and can
bring information out from deep down inside a black hole.

	The pairs of particles coming out of a putative wormhole end
at the black hole
horizon must be different from the pairs produced by the
gravitational field
that lead to the ordinary Hawking radiation, since in the latter,
each particle
is precisely correlated with its corresponding antiparticle, so that
the
positive-energy particles going out actually decrease the information
in the
outside.  The particles escaping to infinity from a wormhole end at
the horizon
would have to be more correlated with other particles going to
infinity than
with their negative-energy partners that go down the hole in order to
increase the information outside.  This correlation with other
particles going
to infinity would need to be brought about by the antiparticles of
these other
particles going to the other end of the wormhole (the end in the high
curvature
region).

	Without knowing how to do any relevant calculations, its
sounds rather
unlikely for everything to work out for wormholes to be able to bring
out {\it
all} the information that goes down an apparent black hole, but maybe
in some
sense any remaining information just persists in the high curvature
region
until it has been brought back out to the outside by wormholes.  A
somewhat
bigger question I have with this is that if the topological changes
of
wormholes
are allowed, why should the wormholes bring all the information back
out to our
universe, rather than dumping some of it in another universe (e.g., a
baby
universe)?  Considerations such as these do lead me to feel that a
loss of
information from our universe may be about as plausible as its
preservation,
but it seems that both are open possibilities.

	For example, instead of wormholes bringing the information
out from
inside black holes, it might be done by tiny threads or tubes or
energy
conduits
which are narrow regions
where the metric and causal structure are much different from that of
the
surrounding spacetime, but which do not change the topology.  If
there were
some principle preventing amplitudes for different topologies, such
as
wormholes, then perhaps information could be prevented from escaping
to other
universes.  However, if amplitudes were allowed for these tubes,
through which
the future timelike direction might be from the high curvature region
near
$r=0$ to just outside what would be the horizon in the absence of
these tubes,
then perhaps, just perhaps, all of the information would always be
brought back
into outgoing radiation as the apparent black hole evaporates away.

\vspace{5 mm}
{\bf 6.  Conclusions}
\vspace{5 mm}

	At an Open Discussion on June 24, 1993, at the Conference on
Quantum
Aspects of Black Holes held at the Institute for Theoretical Physics
of the
University
of California at Santa Barbara, Joe Polchinski asked the conference
participants
to express their opinions about what happens to the information that
falls into
a black hole.  25 voted for option a), ``It's lost.''  39 voted for
b), ``It
comes
out with the Hawking radiation.''  7 voted for c), ``It remains
(accessible) in
a
black hole remnant.''  Finally, 6 voted for d), ``Something else.''

	Of course, this voting for specific options represents a
renormalization of
the varying opinions held within each individual, such as John
Preskill, who
had just previously concluded his talk \cite{PreSB} by saying that in
his
heart
he believed in an $S$-matrix, whereas in his mind he believed in
information
loss.  To illustrate my own varying opinions, permit me to repeat the
subjective
likelihoods I listed at the same conference for the various
possibilities
\cite{PagSB}:
\pagebreak
\begin{tabbing}
C. \= 3. Stable remnant that can't be destroyed space space\=
3\=3\%space space\=3\=5\%\kill
\>{\bf Possibilities for Information}\hspace{25 mm}{\bf Personal
Estimate of
Likelihoods}\\
A. It's lost\>\>30\%\+\\
 1. Evolution forward by $\$$\>\>\>\>9\%\\
 2. Evolution backward by $\$$\>\>\>\>1\%\\
 3. Nonlinear ovolution of $\rho$\>\>\>\>1\%\\
 4. Unpredictable evolution (open system)\>\>\>14\%\\
 5. Inaccessible in a remnant\>\>\>\>5\%\-\\
 B. It comes out with the Hawking radiation\>\>35\%\+\\
 1. Unitary $S$ matrix\>\>\>33\%\\
 2. Nonunitary $S$ matrix\>\>\>\>2\%\-\\
C. It remains accessible in a remnant\>\>\>5\%\+\\
1. Long-lived decaying remnant\>\>\>\>2\%\\
2. Stable remnant that can be destroyed\>\>\>\>2\%\\
3. Stable remnant that can't be destroyed \>\>\>\>1\%\-\\
D. Something else\>\>30\%\+\\
1. Marvelous moment  QM\>\>\>\>5\%\\
2. Gell-Mann--Hartle QM\>\>\>\>5\%\\
3. Other generalized QM\>\>\>10\%\\
4. Overthrow of QM\>\>\>\>5\%\\
5. Other\>\>\>\>5\%\\
\end{tabbing}

	Even in my own mind, these subjective estimates of the
likelihoods are
uncertain by factors of order two (though the sum is not).  I should
also
clarify
that the likelihoods under ``Something else'' were not my estimates
of the
likelihoods that these possibilities are correct, but that they are
correct
{\em and} give results that cannot be listed under the previous
classifications.
Actually, after a bit more reflection, I am now inclined to give
these
possibilities higher likelihoods, so perhaps I should have voted for
d) rather
than b).

	In conclusion, most of the original possibilities
\cite{Pag80} still seem to
be open, with various ones of these amplified or expanded into new
possibilities, such as the various types of remnants that have now
been
suggested.  The introduction of interesting two-dimensional models to
this
problem \cite{CGHS} has opened up a wide range of new attacks, but it
appears
to me that no approach we now know will solve the problem in the
foreseeable
future.  Here is certainly a worthwhile challenge for the
twenty-first century.

\pagebreak
\baselineskip 14.6pt
	{\bf Acknowledgments}:  Appreciation is expressed for the
hospitality of the Aspen Center for Physics in Colorado and of Kip
Thorne and Carolee Winstein at their home in Pasadena, California,
where most of my new calculations reported near the end of this paper
were done.  I was grateful to have the opportunity to report on this
at the California Institute of Technology, the University of
California at Santa Barbara, the Journ\'{e}es Relativistes '93 at the
Universit\'{e} Libre de Bruxelles, the 5th Canadian Conference on
General Relativity and Relativistic Astrophysics, and the Institute
for Theoretical Physics Conference on Quantum Aspects of Black Holes,
where many people gave useful comments.
There and elsewhere I particularly remember discussions with
Peter Aichelburg, Shanta de Alwis, Arley Anderson, Paul Anderson,
Warren Anderson, Abhay Ashtekar, Tom Banks, Julian Barbour,
Andrei Barvinsky, Jacob Bekenstein, Iwo Bialynicki-Birula, Adel
Bilal,
Patrick Brady, Steve Braham, Dieter Brill, R. Brout, David Brown,
Esteban Calzetta, Bruce Campbell, Steve Carlip, Brandon Carter,
Carl Caves, Sidney Coleman, Greg Comer, Ulf Danielsson,
Bahman Darian, Peter D'Eath, Fernando De Felice,
Natalie Deruelle, Venzo De Sabbata, Stanley Deser,
Hector DeVega, Bryce DeWitt, Robbert Dijkgraaf,
Serge Droz, Freeman Dyson, George Ellis, Francois Englert,
Willy Fischler, Faye Flam, Richard Feynman, Jerry Finkelstein,
Tim Folger, Dan Freedman, John Friedman, Valery Frolov,
Juan Garcia-Bellido, Murray Gell-Mann, Bob Geroch,
Gary Gibbons, Steve Giddings, Paul Ginsparg,
Dahlia Goldwirth, Robert Griffiths, David Gross,
Jonathan Halliwell, Petr H\'{a}j\'{\i}\u{c}ek, James Hartle,
Jeff Harvey, Stephen Hawking, Geoff Hayward, Justin Hayward,
Marc Henneaux, Bill Hiscock, Gerard 't Hooft, Jim Horne,
Gary Horowitz, Bei-Lok Hu, Viqar Husain, Chris Isham,
Werner Israel, Ted Jacobson, Renata Kallosh, Nemanja Kaloper,
Claus Kiefer, Tom\'{a}\v{s} Kopf, Pavel Krtou\v{s}, Karel Kucha\v{r},
Gabor Kunstatter, Raymond Laflamme, Kayll Lake,
Peter Landsberg, Kimyeong Lee, T. D. Lee, Andrei Linde,
Seth Lloyd, Jorma Louko, Carlos Lousto, David Lowe,
Elihu Lubkin, Robb Mann, Donald Marolf, Samur Mathur,
Randy McKee, Ray McLenaghan, Aleksandar Mikovi\u{c},
John Moffat, Mike Morris, Sharon Morsink, Valery Mukhanov,
Rob Myers, Dimitri Nanopoulos, William Nelson,
Hermann Nicolai, Igor Novikov, Martin O'Loughlin, Miguel Ortiz,
Cathy Page, Leonard Parker, Yoav Peleg, Roger Penrose,
Asher Peres, Malcolm Perry, Suresh Pillai, Tsvi Piran,
Eric Poisson, Joe Polchinsky, John Preskill, Richard Price,
Jorge Russo, David Salopek, Norma Sanchez, Philippe Spindel,
Andy Strominger, Marcelo Schiffer, Ben Schumacher,
John Schwarz, Dennis Sciama, Jonathan Simon, Lee Smolin,
Rafael Sorkin, Mark Srednicki, John Stewart, Lenny Susskind,
Gary Taubes, L\'{a}rus Thorlacius, Charles Thorn, Kip Thorne,
Claudio Teitelboim, Mike Turner, Sandip Trivedi,
Jason Twamley, John Uglum, Bill Unruh, Erik Verlinde,
Herman Verlinde, Alex Vilenkin, Grisha Vilkovisky, Bob Wald,
Steven Weinberg, John Wheeler, Frank Wilczek, Raymond Willey,
Ed Witten, Bill Wootters, C. N. Yang, Jim York, Tom Zannias,
Yakov Zel'dovich, Zhenjiu Zhang, Wojtek Zurek,
and others, though my memory of the discussions certainly
does not imply no loss of information of their content on my part.
Financial support was provided in part by the Natural Sciences and
Engineering Research Council of Canada.

\end{document}